\documentclass[aps,twocolumn,showpacs,lengthcheck,preprintnumbers]{revtex4-1} 
\usepackage{amsfonts}
\usepackage{amsmath}
\usepackage{amssymb}
\usepackage{graphicx}
\usepackage{fontenc}
\usepackage[normalem]{ulem} 
\usepackage{color}

\usepackage[unicode]{hyperref}
\usepackage{microtype}
\hypersetup{
	pdftitle={two rings},
	pdfauthor={},
	pdfsubject={},
	colorlinks=true,
	linkcolor=blue,
	citecolor=blue,
	filecolor=black,
	urlcolor=blue
}

\newcommand{\omegakd}{\omega_{{k}}^{\smash{(\mathrm{d})}}}
\newcommand{\omegaks}{\omega_{{k}}^{\smash{(\mathrm{s})}}}

\begin{document}
\title{Spinor Bose-Einstein condensates subject to current-density interactions}

\author{Maria Arazo}
\affiliation{
	Departament de F\'isica Qu\`antica i Astrof\'isica, Universitat de Barcelona, 08028 Barcelona, Spain
}
\affiliation{
	Institut de Ci\`encies del Cosmos, Universitat de Barcelona, Mart\'i Franqu\`es 1, 08028 Barcelona, Spain
}
\author{Vicente Delgado}
\affiliation{Departamento de F\'isica, Universidad de La Laguna, Tenerife 38200, Spain}
\author{Montserrat Guilleumas}
\affiliation{
	Departament de F\'isica Qu\`antica i Astrof\'isica, Universitat de Barcelona, 08028 Barcelona, Spain
}
\affiliation{
	Institut de Ci\`encies del Cosmos, Universitat de Barcelona, Mart\'i Franqu\`es 1, 08028 Barcelona, Spain
}
\author{Ricardo Mayol}
\affiliation{
	Departament de F\'isica Qu\`antica i Astrof\'isica, Universitat de Barcelona, 08028 Barcelona, Spain
}
\affiliation{
	Institut de Ci\`encies del Cosmos, Universitat de Barcelona, Mart\'i Franqu\`es 1, 08028 Barcelona, Spain
}
\author{Antonio Mu\~{n}oz Mateo}
\affiliation{Departamento de F\'isica, Universidad de La Laguna, Tenerife 38200, Spain}

\date{\today}
	
\begin{abstract} 
	Recently achieved chiral condensates open intriguing avenues for the study of the chiral properties induced by current-density interactions. An attempt to include these features in a spinor system is presented, which gives rise to a nonlinear, effective spin-orbit coupling that emerges from the differential orbital currents, along with constraints in the conserved quantities due to the linear coupling between spin components.
	Chirality pervades the resulting spectrum of stationary states and their dynamical stability, which are explored in plane waves,  dark and bright solitons, and Josephson vortices. Our analytical and numerical results reveal the destabilizing role of polarization and Josephson currents, and support the existence of stable nonlinear states built of linear superpositions of plane waves.
\end{abstract}


\maketitle
	

\section{Introduction}
Ultracold Bose-condensed gases of atomic species subject to interactions that are proportional to the local atomic current-density have been recently achieved~\cite{Frolian2022}. 
The road to realization involves electrically-neutral matter systems with pseudo-spin $1/2$ coupled to laser fields that give rise to synthetic electromagnetism; the emergent gauge fields turn out to be the effect of geometrical phases accumulated in the adiabatic path of the optically dressed atomic states (see~\cite{dalibard2011,Edmonds2013} and references therein). 
The resulting systems exhibit chiral properties~\cite{Chisholm2022} when they are restricted to their lowest energy bands and are governed by an effective Hamiltonian that includes a current-density term and operates on a scalar order parameter.
The theoretical model can be mapped into the 1D reduction of a 2D topological gauge theory that breaks Galilean invariance, and was predicted to host the chiral solitons~\cite{Aglietti1996,Jackiw1997,Griguolo1998} that have been observed in the experiment~\cite{Frolian2022}. 
Further features of this chiral scalar theory have been addressed in the last years~\cite{Edmonds2013,Edmonds2013b,Edmonds2015,Dingwall2018,Dingwall2019,Bhat2021,Jia2022,Jintao2023}.

In this work, we 
aim to implement a spinor system from the effective chiral scalar condensate in order to explore a long-Josephson bosonic junction with chiral properties.
To this end, we 
model an effective two-component spinor Bose-Einstein condensate (BEC) 
as realized in regular extended bosonic junctions~\cite{Schweigler2017,Pigneur2018}.
In this setting, we study the Josephson dynamics of extended chiral states as a generalization of a chiral point-like Josephson junction~\cite{Edmonds2013b}.
Although one can still discuss the state properties with respect to the population imbalance between components (or spin polarization), the relative (or spin) current density $J_s$ emerges as a key quantity: it is linked to the Josephson current that flips the (pseudo-) spin through a continuity equation, and causes distinct interaction properties between particles with different spin. 
The spin current replicates also the dynamical response of a superconducting, chiral junction to the action of an external magnetic field.
Then, as in weak superconducting systems, Josephson vortices (localized loop currents between spin components) with cores at the junction are generated.
	
The paper is structured as follows. 
In Sec.~\ref{sec:Model}, we present the equations of motion and relevant quantities, paying attention to particular features of the ring geometry under consideration, and classify the different types of stationary states that can be found.
In Sec.~\ref{sec:pw}, we analyze plane waves and their superpositions, whereas Sec.~\ref{sec:Solitons} focuses on solitonic states. Finally, in Sec.~\ref{sec:Conclusions}, we present our conclusions.

	
\section{Model}\label{sec:Model}
We consider a linearly coupled, two-component (labeled $\uparrow$ and $\downarrow$) BEC in a rotating ring geometry of radius $R$ and subject to the same (intra-component) current-density interaction of non-dimensional strength $\kappa_\uparrow=\kappa_\downarrow=\kappa$. We assume $\kappa>0$ without loss of generality. In order to see the effect of this type of interaction more clearly, we also assume that there is no contact interparticle interaction in the system, $g_{\uparrow\uparrow}=g_{\downarrow\downarrow}=g_{\uparrow\downarrow}=0$. The equation of motion is modeled by the mean-field Gross-Pitaevskii-like (GP) equation for the pseudo-spin-$1/2$ wave function $\Psi=[\psi_\uparrow \;\psi_\downarrow]^T$
\begin{align}
	i\hbar\frac{\partial\Psi}{\partial t}  =
	\left(\begin{array}{cc} 
      \displaystyle
		\frac{\hat\Pi^2}{2M}+\kappa \hbar J_{\uparrow}& -\nu \\
		-\nu & 
      \displaystyle
            \frac{\hat\Pi^2}{2M}+\kappa \hbar J_{\downarrow}\end{array} \right) \Psi,
	\label{eq:gp}
\end{align}
where $M$ is the particle mass, $\hat\Pi=\hat p-M\Omega R$ is the mechanical momentum operator in the frame rotating with angular velocity $\Omega$, $\hat p=-i\hbar\partial_x$ is the canonical momentum operator,  $\nu>0$ is the energy of the linear coupling, and $J_\sigma =\hbar(\psi_\sigma^*\partial_x\psi_\sigma-\psi_\sigma\partial_x\psi_\sigma^*)/(i2M)$ are the component current densities measured with respect to the laboratory frame (with $\sigma=\uparrow,\downarrow$).

The number of particles in each component is not conserved due to the linear coupling, but the total number of particles in the ring $N=\oint dx \Psi^\dagger\Psi=\oint dx\,(n_\uparrow+n_\downarrow)$, where $n_\sigma=|\psi_\sigma|^2$, is a conserved quantity by means of the preserved $U(1)$ symmetry \cite{Son2002}. As a result,
the system satisfies the continuity equation 
\begin{align}
\partial_t n+\partial_x (J-n\Omega R)=0,
\label{eq:cont}
\end{align}
where $n(x,t)=n_\uparrow+n_\downarrow$  and $J(x,t)=J_\uparrow+J_\downarrow$ stand for the total particle and current densities, respectively. In addition, the local population imbalance between components or spin density, $n_s(x,t)=n_\uparrow-n_\downarrow$, fulfills a second continuity equation,
\begin{align}
\partial_t n_s+\partial_x (J_s-n_s\Omega R)=\mathcal{I}_\varphi,
\label{eq:cont1}
\end{align}
where $J_s=J_\uparrow-J_\downarrow$ is the spin current density, i.e., the relative current between components. The source term in the right-hand side of Eq.~\eqref{eq:cont1}, $\mathcal{I}_\varphi=({4\nu}/{\hbar})\sqrt{n_\uparrow n_\downarrow}\,\sin\varphi$, where $\varphi=\arg\psi_\uparrow-\arg\psi_\downarrow$ is the relative phase, can be identified as twice the Josephson current flowing between components~\cite{Barone}.

	The stationary states $\Psi(x,t)=\Psi(x)\,\exp(-i\mu\,t/\hbar)$, with energy eigenvalue $\mu$, satisfy the time-independent equation $\hat H\Psi=\mu\Psi$, where $\hat H$ is the nonlinear Hamiltonian matrix in Eq.~\eqref{eq:gp}, and also the time-independent versions of Eqs.~\eqref{eq:cont} and \eqref{eq:cont1}. For them,  $\mathcal{J}=J-n\,\Omega R $ is always the constant total current in the rotating frame while, from $\partial_x (J_s-n_s\,\Omega R)=\mathcal{I}_\varphi$, one can find a constant spin current density, $\mathcal{J}_s=J_s-n_s\,\Omega R $, only when $\varphi=j\,\pi$ (for $j$ integer).

As in the equivalent scalar condensate \cite{Arazo2023}, one can define a corresponding average energy,
\begin{eqnarray}
	E=\oint dx \left(  \sum_\sigma \frac{\psi_\sigma^\ast\hat{\Pi}^2 \psi_\sigma}{2M}   
	- 2\nu\, \sqrt{n_\uparrow n_\downarrow}\,\cos\varphi \right),
	\label{eq:energy}
\end{eqnarray}
which does not explicitly depend on $\kappa$ and includes the Josephson (coupling) energy
$E_\varphi=-2\,\nu \,\mathrm{Re}(\psi_\uparrow^\ast\psi_\downarrow)=  - 2\nu\, \sqrt{n_\uparrow n_\downarrow}\,\cos\varphi$.
However, differently to the particle number conservation,
it turns out that $E$ is not a conserved quantity.
This can be seen from the Ehrenfest's theorem applied to the Hamiltonian matrix $\hat H$  in Eq. (\ref{eq:gp}) [rewritten below in Eq. (\ref{eq:Hamiltonian})], which states the equality $ d/dt \langle \hat H \rangle=\langle \partial \hat H /\partial t \rangle$ between expectation values and gives
\begin{align}
\frac{d}{dt} E+{\kappa\hbar}\oint dx\,(J_\uparrow\,\partial_t n_\uparrow +J_\downarrow\,\partial_t n_\downarrow)=0.
\label{eq:Efail}
\end{align}
The second term acts as an energy source and prevents $E$ from being, in general, a conserved quantity. Despite this fact, we are able to find stationary states that are dynamically stable against small perturbations. In this regard, the present system is not dissimilar to other dissipative non-linear systems, as, for instance, long-lived Bose-Einstein condensates of exciton-polaritons (see, e.g., \cite{Bloch2022} and references therein), where the analysis focuses on the generation and stability of steady-state configurations.

\subsection{Linear excitations}
	
The linear excitations of stationary states $\Psi(x,t)$ can be obtained by solving the Bogoliubov equations~\cite{Pitaevskii}. 
They result from introducing the perturbed state 
\begin{align}
	\phi_\sigma(x,t)&= e^{-i\mu t/\hbar}\,
      \bigg\lbrace \psi_\sigma(x)\,+ \nonumber \\ &  
	\sum_j \left[ u_{j \sigma}(x)\,e^{-i\omega_j t}+v^*_{j \sigma}(x)\,e^{i\omega_j^* t} \right] \bigg\rbrace 
\end{align} 
into the GP Eq.~\eqref{eq:gp}, where $j$ indexes the linear modes and $\sigma$ the condensate component. The vector of the excitation-mode amplitudes, $\delta\psi_j=[u_{j \uparrow}\;v_{j \uparrow}\;u_{j \downarrow}\;v_{j \downarrow}]^T$, solves the linear system of equations 
\begin{equation}
	\begin{pmatrix}
		\hat B_\uparrow & -\nu\,\sigma_z \\
		-\nu\,\sigma_z & \hat B_\downarrow
	\end{pmatrix}
	\delta\psi_{j}
	= \hbar\omega_j\,\delta\psi_{j},
\end{equation}
with the $2\times 2$ Bogoliubov operators,	
\begin{equation}
	\hat{B}_\sigma =
	\begin{pmatrix}
            \displaystyle
		\hat H_ \sigma +\frac{\hbar\kappa}{2M}{\psi_ \sigma\,\mathcal{C}(\psi_ \sigma^\ast,\,\hat{p})} &
            \displaystyle
		-\frac{\hbar\kappa}{2M}{\psi_ \sigma\,\mathcal{C}(\psi_ \sigma,\,\hat{p})} \\
            \displaystyle
		-\frac{\hbar\kappa}{2M}{\psi_ \sigma^\ast\,\mathcal{C}(\psi_ \sigma^\ast\,\hat{p})} &
            \displaystyle
		-\hat H_ \sigma^\ast +\frac{\hbar\kappa}{2M}{\psi_ \sigma^\ast\mathcal{C}(\psi_ \sigma\,\hat{p})}
	\end{pmatrix}\,,
\end{equation}
where $\hat{H}_\sigma={\hat{\Pi}^2}/{2M}+\kappa\hbar J_\sigma-\mu$, and we have introduced the operator $\mathcal{C}(\psi_\sigma,\hat p)=\psi_\sigma\,\hat p-(\hat p \psi_\sigma)$ for short notation. Pure real modes $\delta \psi_j$ are stable excitations, but complex modes signal dynamical instabilities.

\subsection{Constraint of the ring geometry}

Both the ring geometry and the spinor nature introduce some constraints in the theory of the system under consideration that are worth emphasizing. 
It has been demonstrated that, in open geometries, the scalar chiral model can be mapped into a (reduced) topological gauge theory by means of the nonlocal (Jordan-Wigner-like) transformation  \cite{Aglietti1996}
\begin{align}
\psi'(x,t)=\psi(x,t)\,\exp\left(-i\frac{\kappa}{2}\int^x_{x_0} dy\,|\psi(y,t)|^2\right),
\label{eq:WJ}
\end{align}
by means of which the kinetic momentum acquires (in addition to the canonical momentum) a density-dependent contribution $p_n=-\hbar(\kappa/2)|\psi'|^2$.
The advantage of the resulting theory resides in the existence of a local Lagrangian and the subsequent definition of conserved quantities~\cite{Jackiw1997}.
Unfortunately, the ring geometry can frustrate the mapping between theories. Since the wave function is single valued in both theories, and the density profile is preserved by Eq.~\eqref{eq:WJ}, the phase $\theta=\arg\psi$ (and $\theta'=\arg\psi'$) is restricted to jump in integer multiples of $2\pi$ when it winds around the ring, $\Delta\theta=\theta(2\pi R)-\theta(0)$. 
Therefore, from Eq.~\eqref{eq:WJ}, the mapping is possible only if $\Delta\theta'=\Delta\theta+\kappa N/2=2\pi j^\prime $, hence only if $\kappa N/2=2\pi j $, for $j$ and $j^\prime$ integers. 
In other words, the transformation given by Eq.~\eqref{eq:WJ} is allowed in the ring only for quantized values of the total number of particles,
\begin{align}
{N}=\frac{4\pi}{\kappa}\,j \,.
\label{eq:quantaN}
\end{align}
The lack of mapping has interesting consequences on the scalar chiral systems. For instance, the  transition between ground states $\smash{\psi^\prime_{q^\prime}=\sqrt{N/(2\pi R)}\exp(iq^\prime x)}$ with different winding number $q^\prime$  found in Ref.~\cite{Edmonds2013} for varying number of particles $N$ is not seen within the theory with current-density interactions. Here, $q$ (then $q=q^\prime+\kappa N/2$ when the mapping exits) is just the wave number of the mechanical momentum, independent of the number of particles, and $\psi_0$ (with $q=0$) is always the ground state.

The spinor character of the system brings further constraints in the search of local Lagrangians and conserved quantities, as reflected by Eq. (\ref{eq:Efail}). A mapping between spinor systems equivalent to Eq.~\eqref{eq:WJ} would involve two parallel transformations for the two spin components, in a similar way as Wigner-Jordan transformations operate in discrete ladders (see, for example,~\cite{Hur2017}).
But in this case, new non-local phases are expected to emerge in the linear-coupling term.

\subsection{Types of solutions}

It is useful to write the nonlinear Hamiltonian~\eqref{eq:gp} in terms of the Pauli matrices $\boldsymbol{\sigma}=(\sigma_x,\sigma_y,\sigma_z)$ and the $2\times 2$ identity matrix $I_2$ as 
\begin{align}
	\hat H=\left(\frac{\hat{\Pi}^2}{2M}  + \frac{\kappa\hbar J}{2} \right) I_2 +\frac{\kappa\hbar J_s}{2}\,\sigma_z-\nu\,\sigma_x\,.
	\label{eq:Hamiltonian}
\end{align}
Hence, one can identify an effective spin-orbit-coupling term $(\kappa\hbar J_s/2)\,\sigma_z$, which shifts the energies of the spin components according to the axial (orbital) spin current $J_s$. 
There are two types of stationary states depending on the absence or presence of spin current.
For the former type ($J_s=0$), the Pauli matrix $\sigma_x$ commutes with the Hamiltonian, so one can find common eigenstates that satisfy
$\psi_\downarrow(x,t) =\pm \psi_\uparrow(x,t)$, which transforms the coupled equations~\eqref{eq:gp} into the single equation
\begin{align}
	i\hbar\frac{\partial\psi_\uparrow}{\partial t}= 	\left(\frac{\hat\Pi^2}{2M}+\kappa \hbar J_{\uparrow} \mp \nu \right)\psi_\uparrow.
	\label{eq;single_gp}
\end{align}
In this case, one recovers all the stationary states known for a single-component condensate~\cite{Arazo2023}:
\begin{equation}
	\psi_\uparrow(x,t)=\sqrt{\alpha+\beta\,{\rm dn}^2(x/\xi,\mathfrak{m})}\;e^{i\theta_\uparrow(x)-i\mu t/\hbar},
	\label{eq:dn_general}
\end{equation}
with phase 
\begin{equation}
	\theta_\uparrow(x)=k_{\tiny \Omega}{x} +\frac{M\xi\mathcal{J}}{2\hbar(\alpha+\beta)}\Pi(\eta;x/\xi,\mathfrak{m}),
	\label{eq:dnPhase}
\end{equation}
where we have used the Jacobi ${\rm dn}$ function with parameter $\mathfrak{m}\in[0,1]$ and characteristic length $\xi={\hbar}/{\sqrt{ M \kappa\hbar|\Omega \beta| R}}$,
and the incomplete elliptic integral of the third kind $\Pi(\eta;x/\xi,\mathfrak{m})$, with $\eta=\mathfrak{m}\beta/(\alpha+\beta)$~\cite{Abramowitz}. The real parameters $\{\mathfrak{m},\,\alpha,\,\beta\}$ are self-consistently determined for a particular system with $\{R,\,N,\,\Omega\}$ (see Ref.~\cite{Arazo2023} for details). The energy eigenvalues are shifted by $\mp \nu$ with respect to the scalar case:
\begin{equation}
	\mu=\left(\mathfrak{m}-2-3\frac{\alpha}{\beta}\right)\frac{\hbar^2}{2M\xi^2}+\frac{\mathcal{\kappa\hbar\,J}}{2}\mp \nu\,.
	\label{eq:dn_mu}
\end{equation}
	
For states with $J_s\neq 0$, in the general case, one has to deal with the two coupled equations~\eqref{eq:gp}, and a double degeneracy in the energy eigenvalue is obtained for a given $|J_s|$ according to the signs of the spin current. 
Still, in the limit of $\nu\rightarrow 0$, the symmetry with $\sigma_z$ (which results from $[\sigma_z,\hat H]=0$) is slightly broken by a tiny $\nu\neq 0$ and one could expect the stationary states to approach the eigenstates of $\sigma_z$, with $\psi_\sigma/\psi_{\bar\sigma}\approx 0$, thus realizing a population selftrapping. As we discuss later, this is indeed the scenario shown by our results, and the relevant parameter that controls these regimes is the ratio $\gamma=2\nu/(\kappa\hbar J)$ between the linear coupling and the interaction.

From now on, we introduce the average number density $n_0=N/(2\pi R)$, the rotation wave number $k_{\tiny \Omega}=M\Omega R/\hbar$, and the normalized interaction strength $\tilde \kappa=\kappa N/(2\pi)$. In addition, we make use of the ring units $\{R,\,\Omega_0^{-1}=M R^2/\hbar, \,\hbar\Omega_0\}$ as length, time, and energy units to write non-dimensional quantities, which we denote by tildes, e.g., $\tilde \Omega=\Omega/\Omega_0$.
	
\section{Plane wave spectrum} \label{sec:pw}
	
As in the corresponding scalar system, the equation of motion~\eqref{eq:gp} is translational invariant and can be solved by plane-wave states
\begin{equation}
	\psi_q(x,t)=\binom{\sqrt{n_{0\uparrow}}}{\pm\sqrt{n_{0\downarrow}}}\,e^{i(q\,x-\mu_q t/\hbar)}\,,
	\label{eq:pw}
\end{equation}
where $n_{0\sigma}$ are constant densities and the $\pm$ sign accounts for the relative phase between components, $\varphi=0$ or $\pi$, respectively. 
The common wave number $q$ is quantized in the ring, $qR=0,\pm 1,\pm2,\dots$, and the total and spin current densities in the laboratory frame are $ J_q=\hbar q\,n_0/M$ and $J_{qs}=\hbar q\, n_{0s}/M$.  
	
The absence ($J_{qs}=0$) or presence ($J_{qs}\neq 0$) of spin current corresponds to the absence or presence, respectively, of local population imbalance between components or polarization, given by $n_{0s}$. 
The first kind ($n_{0s}=0$) corresponds to equal spin populations, $n_{0\uparrow}=n_{0\downarrow}=n_0/2$, and has energy eigenvalues
\begin{equation}
	\mu_q^{(\mp)}=\frac{(\hbar q-M\Omega R)^2}{2M}+\kappa\frac{\hbar    J_q}{2}\mp \nu\,,
	\label{eq:muPW1}
\end{equation} 
which describe two separated branches of the dispersion. These states are particular cases of the general solution~\eqref{eq:dn_general} when $\mathfrak{m}=0$.

On the other hand, the polarized states ($n_{0s}\neq 0$) have spin current $\smash{J_{qs}= \pm J_q \sqrt{1- {{\gamma_q}^2}}}$, with $ \gamma_q=2\nu/(\kappa\hbar J_q)$, and give rise to two overlapping energy branches for opposite signs of $J_{qs}$:
\begin{equation}
	\mu^{(s)}_q=\frac{(\hbar q-M\Omega R)^2}{2M}+\kappa\hbar    J_q\,.
	\label{eq:muPW2}
\end{equation} 
But this overlap is not trivial: positive- (negative-) $q$ states exist only in the out-of- (in-) phase branch of the dispersion relations.  
The parameter $| \gamma_q|=|4\pi M R \nu/(\kappa N\hbar^2 q) |$, which reflects the ratio between the linear coupling and the interaction energy terms in the equation of motion, marks the transition between both types of plane waves. While  the unpolarized states exist for arbitrary values of $ \gamma_q$, the polarized ones exist only for high interactions $| \gamma_q|< 1$, thus for $q\neq 0$ and $|q|>|2 M  \nu/(\kappa \hbar^2 n_0)|$. They can be thought of as nonlinear bifurcations of the unpolarized states at $| \gamma_q|=1$.

\begin{figure}[tb]
      \includegraphics[width=\linewidth]{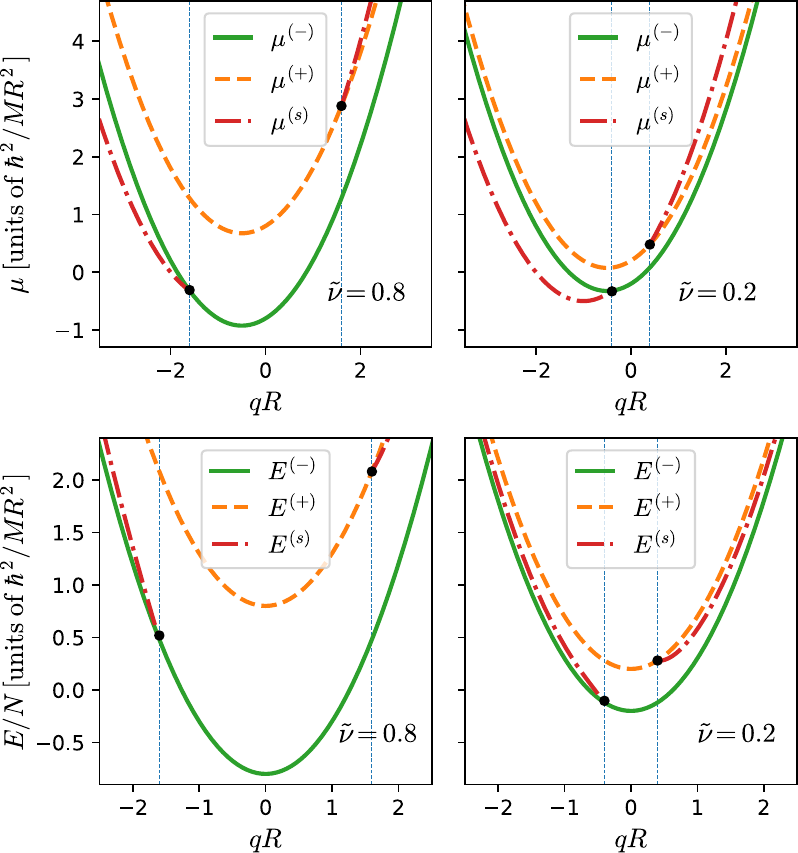}
	\caption{Energy eigenvalue (top) and energy per particle (bottom) of stationary plane waves with wave number $q$ for two values of the linear coupling $\nu$ at $\Omega=0$. The solid circles indicate the bifurcation points of states with non-vanishing spin current density  $J_s$ (thick dot-dashed lines), so that they do not exist within the region limited by the vertical dashed lines. In a ring trap of radius $R$, only the states with $qR=0,\pm 1,\pm 2,\dots$ are allowed, and the dispersion curves are restricted to a discrete set of points.} 
	\label{fig:muE}
\end{figure}
 
Figure~\ref{fig:muE}~(top panel) shows the energy eigenvalues $\mu_q^{\smash{(\mp)}}$ and $\mu_q^{\smash{(s)}}$ of plane waves in the absence of rotation, given respectively by Eqs.~\eqref{eq:muPW1} and \eqref{eq:muPW2}.
Despite the fact that the eigenvalue of states with non-vanishing spin current density ${\mu_q^{\smash{(s)}}}$ becomes the lowest (for negative wave numbers) when $\nu<(\kappa\hbar \,n_0)^2/(2M)$, the average energy~\eqref{eq:energy},
\begin{align}
	\frac{E_q^{(\mp)}}{N}&=\frac{(\hbar q - M \Omega R)^2}{2M} \mp \nu, \nonumber \\
	\frac{E_{q}^{(s)}}{N}&=\frac{(\hbar q - M \Omega R)^2}{2M} \mp \nu |\gamma_q|\,,
	\label{eq:Eq}
\end{align}
does not (see bottom panel).
Hence, the polarized states, when they exist at $|\gamma_q|<1$, are not the ground states of the system according to the average energy Eq. (\ref{eq:energy}).
In contrast, spinor systems with contact interparticle interactions present ground states with either balanced or imbalanced (i.e., spin polarized) populations according to the particular values of the interaction strengths~\cite{abad2013}.

\subsection{Dynamical stability of plane waves}

\begin{figure}[tb]
	\includegraphics[width=\linewidth]{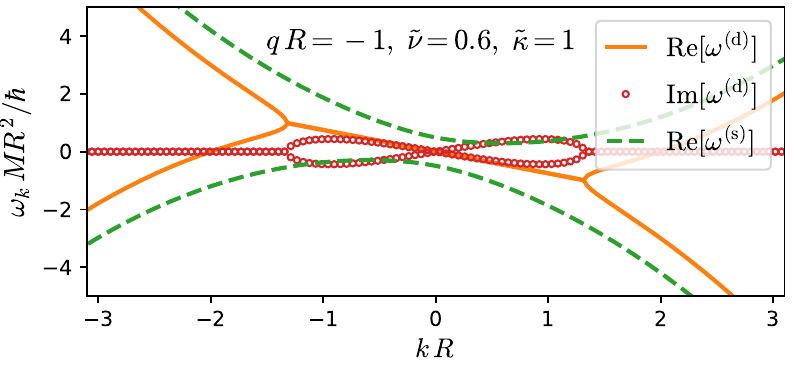}\\ 
	\includegraphics[width=\linewidth]{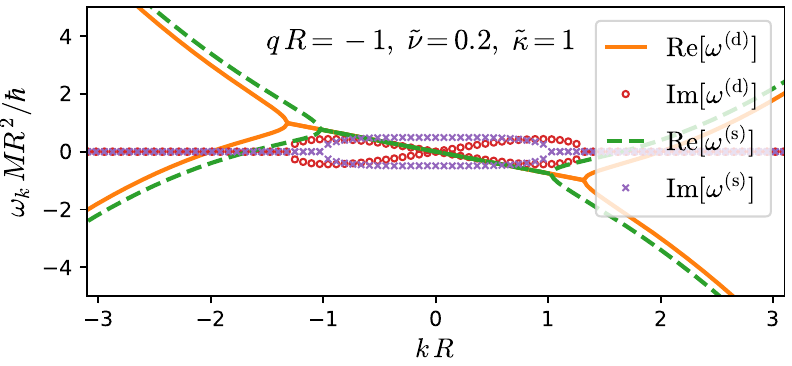}
	\caption{Frequency of linear excitations of unpolarized plane waves with wave number $qR=-1$ and interaction parameter $\kappa N=2\pi$ ($\tilde{\kappa}=1$). Unstable density modes appear for any value of the linear coupling. Spin instabilities appear also at low values of the linear coupling (bottom), while they are suppressed at a higher coupling (top). }
	\label{fig:w_n1}
\end{figure}
\begin{figure}[tb]
	\includegraphics[width=\linewidth]{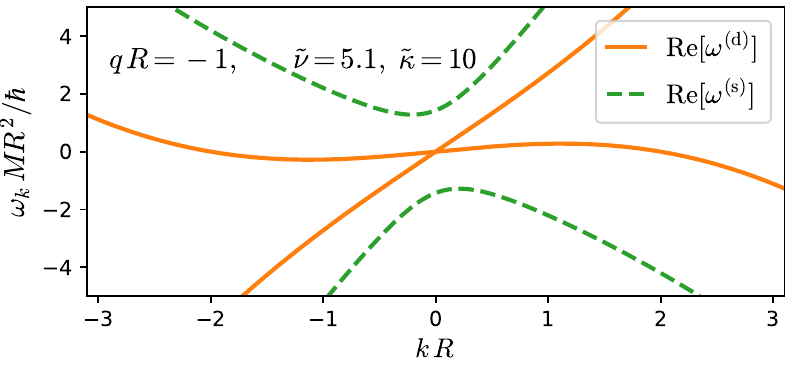}\\ 
	\includegraphics[width=\linewidth]{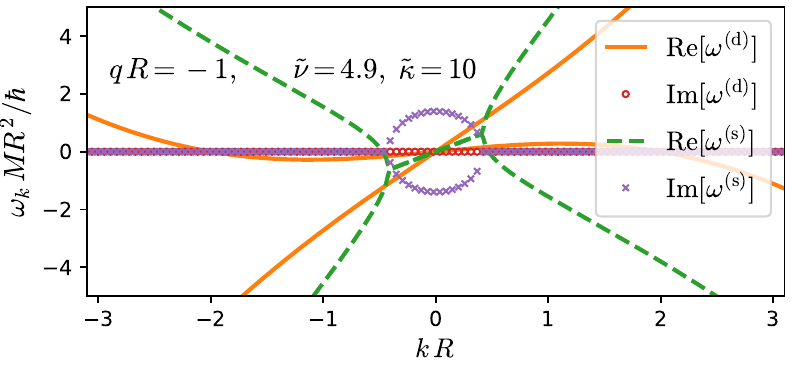}
	\caption{Frequency of linear excitations of unpolarized plane waves with wave number $qR=-1$ and interaction parameter $\kappa N=20\pi$ ($\tilde{\kappa}=10$). The high density (see Fig.~\ref{fig:w_n1} for comparison) suppresses the unstable density modes (top), but unstable spin modes still appear at a low coupling (bottom).}
	\label{fig:w_n10}
\end{figure}

For plane waves with wave number $q$ in the absence of spin currents ($J_s=0$ and $n_{0\uparrow}=n_{0\downarrow}=n_0/2$), one can find linear excitations with equal phase (or density modes), $u_{j\uparrow}=u_{j\downarrow}\equiv u_{j}$ and $v_{j\uparrow}=v_{j\downarrow}\equiv v_{j}$, and out-of-phase excitations (or spin modes), $u_{j\uparrow}=-u_{j\downarrow}\equiv u_{j}$ and $v_{j\uparrow}=-v_{j\downarrow}\equiv v_{j}$. 
Their dispersion reads, respectively,
\begin{equation}
	\label{eq:wd}
	\hbar\omegakd=\frac{\hbar^2 k}{M}\left( q + \frac{\kappa n_0}{4} - k_{\tiny \Omega} 
	\pm \frac{1}{2}\sqrt{k^2 +\frac{M\omega_q}{\hbar}    }\,\right)\,,
\end{equation}
\begin{align}
	\label{eq:ws}
	\hbar\omegaks &= \frac{\hbar^2 k}{M}\left( q + \frac{\kappa n_0}{4} - k_{\tiny \Omega}\right)\pm  \nonumber\\ &
	\sqrt{\frac{\hbar^2 k^2}{2M}\left(\frac{\hbar^2k^2}{2M}+\frac{\hbar\omega_q}{2} +{4 \nu}\right) + 4\nu^2\left(1+\frac{1}{ \gamma_q}\right) } \,, 
\end{align}
where we have introduced the energy term $\hbar \omega_q={(\hbar\kappa n_0)^2}/{4M} + {2\kappa\hbar J_q}$.
The dispersion of the density modes $\omegakd$~\eqref{eq:wd} does not depend explicitly on the coherent coupling $\nu$ and reproduces the linear excitation of single-component condensates~\cite{Arazo2023}: for low wave numbers, $\omegakd$ is linear in $k$ and tends to zero in the $k\to 0$ limit. 
On the other hand, the dispersion of spin modes $\omegaks$~\eqref{eq:ws} shows an energy gap due to the presence of the coherent coupling $\nu$. 
The gap appears (in general, not at $k=0$, but) at the wave number that solves $\partial_k \omegaks=0$. 
	
The spectrum associated with Eqs.~\eqref{eq:wd} and \eqref{eq:ws} contains unstable modes when $\omega_q<0$ or $\gamma_q<0$; both types of instabilities appear only for negative wave numbers. 
The condition $\omega_q<0$, as in scalar condensates, produces modulation instabilities of the total density when $q<-\kappa n_0/8$. 
On the other hand, the condition for spin-density instabilities, $\gamma_q<0$, is the same as for the existence of polarized plane waves when $|\gamma_q|<1$, whose bifurcation point occurs at $\gamma_q=-1$.
This value indicates also the first crossing of the two dispersion branches~[\eqref{eq:wd} and \eqref{eq:ws}] at $k=0$, whereas further crossings take place in the dispersion of unstable states (for $\gamma_q>-1$) at $\hbar k=\pm\sqrt{-2M\nu\,(2\nu+\kappa\hbar J_q)}$. 
The fact that the emergence of spin-density instabilities is associated with the existence of polarized plane waves seems to point to the potential stability of the latter, which is an intriguing feature, since, according to Eq. (\ref{eq:energy}), these states have higher energy that the unpolarized plane waves (see Fig. \ref{fig:muE}). As we show later, although of different type, instabilities are also present in polarized plane waves.
	
	\begin{figure}[tb]
		\includegraphics[width=\linewidth]{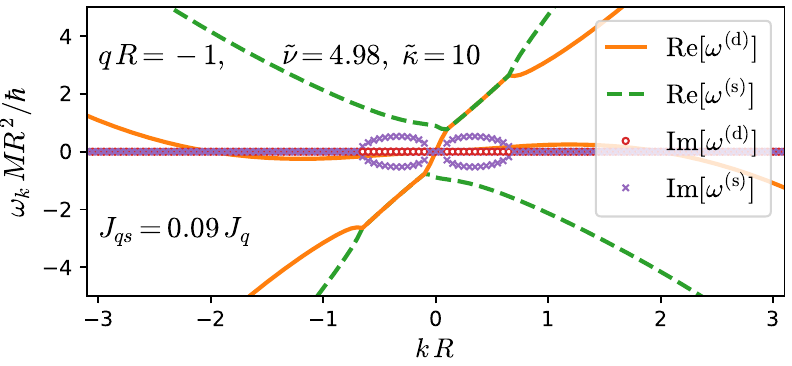}\\
		\includegraphics[width=\linewidth]{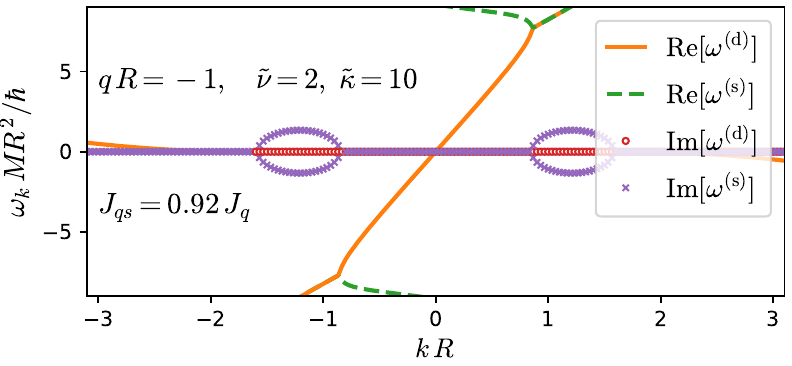}
		\caption{Frequency of linear excitations of polarized plane waves with wave number $qR=-1$ and interaction parameter $\kappa N=20\pi$ ($\tilde{\kappa}=10$). Both density-like and spin-like modes are unstable for any value of the linear coupling. However, the instabilities now arise from the collision of different excitation branches. }
		\label{fig:w_p_n10}
	\end{figure}
Figure \ref{fig:w_n1} shows the linear excitations of unpolarized plane waves with $qR=-1$ and $\tilde\kappa=1$ for two values of the coherent coupling. 
At high coupling $\tilde\nu=0.6$ (top panel), the unstable spin modes are suppressed ($\gamma_q<-1$), but there are unstable density modes ($\omega_q<0$), as indicated by the existence of complex frequencies $\mbox{Im}[\omega^{\smash{(d)}}]\neq 0$ (open circles). 
For $\tilde\nu=0.2$ (bottom panel), both types of instabilities occur, since $\gamma_q=-0.4$ and  $\omega_q=-1.75\,\hbar/(MR^2)$ (the same as for $\tilde\nu=0.6$). 
Due to the constant energy term ${(\hbar\kappa n_0)^2}/{4M}$ in $\hbar\omega_q$, high densities can suppress the unstable density modes. This is shown in Fig. \ref{fig:w_n10}, which depicts the dispersion of plane waves with $qR=-1$ and $\tilde\kappa=10$. 
However, unstable spin modes can still appear if the coherent coupling is not high enough, as shown in the bottom panel at $\tilde\nu=4.9$.

States with spin currents ($J_s\neq 0$) present notable differences in the dispersion of linear excitations. Now, the splitting between total-density and spin-density branches is not meaningful in general cases.  Yet, for negative wave numbers, low number densities trigger  instabilities that closely  resemble those of the total-density modes in unpolarized states (as in Fig. \ref{fig:w_n1}), while those of the spin-density modes are suppressed. 
The distinctive feature appears at higher densities, and it is related to instabilities (for $q<0$) produced by the collision of excitation branches at $k\neq0$, as can be seen in Fig. \ref{fig:w_p_n10}.

\subsection{Plane-wave superpositions at $\Omega=0$}
	
\begin{figure}[tb]
	\includegraphics[width=\linewidth]{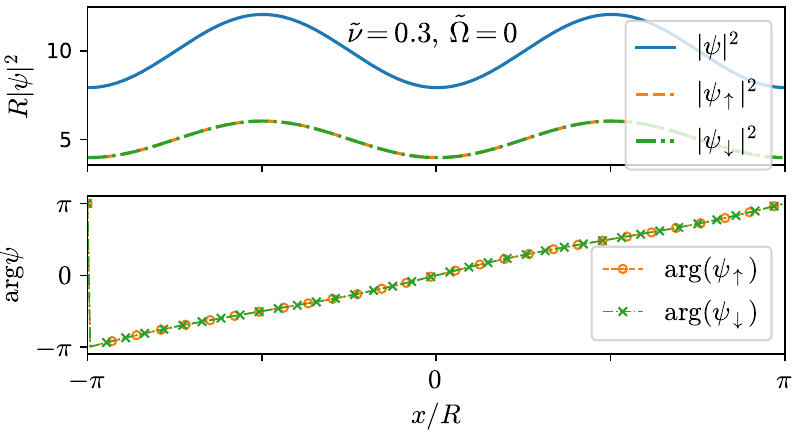}
	\includegraphics[width=\linewidth]{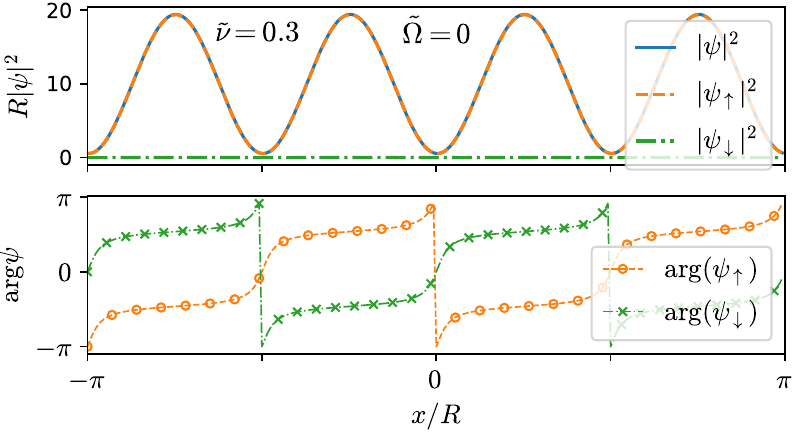}
	\caption{Linear superpositions of plane waves~\eqref{eq:complex} with linear coupling $\tilde\nu=0.3$ in the absence of rotation for an unpolarized state with wave number $qR=1$ and vanishing spin current $J_s=0$ (top), and a polarized state with $qR=2$ and $J_s\neq 0$ (bottom).}
	\label{fig:complex}
\end{figure}
\begin{figure}[tb]
	\includegraphics[width=\linewidth]{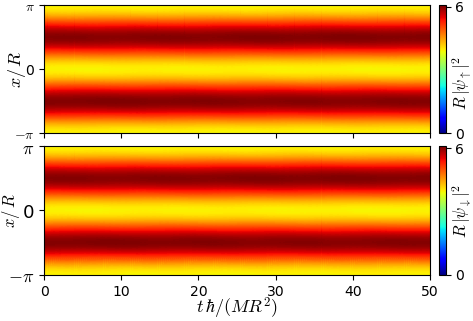}
	\caption{Real-time evolution of the unpolarized linear superpositons shown in Fig. \ref{fig:complex}(a), which is stable. We solve the equation of motion~\eqref{eq:gp} adding small-amplitude sinusoidal perturbations to the initial state.}
	\label{fig:evol_linear}
\end{figure}
	
Although Eq.~\eqref{eq:gp} is a nonlinear equation, it becomes linear if $J_\sigma=0$, and then it admits the same solutions as the free Schr\"odinger equation for a spinor; the usual standing waves $\sin(qx)$ and $\cos(qx)$ fulfill this condition at $\Omega=0$. 
Thus, as one can expect, linear superpositions of these waves with real coefficients are also stationary states~\cite{Arazo2023}. 
Interestingly, complex superpositions with nonzero current,
\begin{align}
	\psi_{\alpha q}=
    \binom{\,\sqrt{n_{\alpha\uparrow}}}{\pm\sqrt{n_{\alpha\downarrow}}}
    \left[  \frac{1-\alpha}{2}\,e^{-iqx} + \frac{1+\alpha}{2}\,e^{i(qx+\phi)}  \right]e^{-i\mu_{\alpha q} t/\hbar}\,,
	\label{eq:complex}
\end{align}
where $\alpha$ and $\phi$ are real numbers, solve Eq.~\eqref{eq:gp} as well. 
These superpositions have constant current densities $J_\sigma=\alpha n_{\alpha\sigma}\hbar q/M$ and $ J= J_q\, 2\alpha/(1+\alpha^2) $; from normalization, one finds $n_\alpha=n_{\alpha\uparrow}+n_{\alpha\downarrow}=2\,n_0/(1+\alpha^2)$. 
As before, solutions of the type of Eq.~\eqref{eq:complex} with both $J_s=0$ and $J_s\neq 0$ are possible. 
In the former case, the energy eigenvalue and average energy per particle read
\begin{align}
	\mu_{\alpha q}^{(\mp)}=\frac{\hbar^2 q^2}{2M}+\frac{\alpha}{1+\alpha^2}\,\kappa\,{\hbar J_q} \mp \nu,
	\label{eq:mua1}
	\\
	\frac	{E_{\alpha q}^{(\mp)}}{N}= \frac{\hbar^2q^2}{2M} \mp \frac{2\,\nu}{1+\alpha^2},
\label{eq:Ea1}
\end{align}
and there is no restriction for $\alpha$ other than normalization.

For $J_s\neq 0$, there exists a population imbalance given by  $n_{\alpha s}=n_{\alpha\uparrow}-n_{\alpha\downarrow}=\pm n_0 \sqrt{[2/(1+\alpha^2)]^2-( \gamma_q/\alpha)^2}$, which constraints $\alpha\neq 0$ to be in the interval $|\alpha| \in[1-\sqrt{1- {\gamma_q}^2},\,1+\sqrt{1- {\gamma_q}^2}\,]/| \gamma_q|$. 
The energy eigenvalue and the average energy become
\begin{align}
	\mu_{\alpha q}^{(s)}=\frac{\hbar^2 q^2}{2M}+\frac{2\alpha}{1+\alpha^2}\,\kappa\,{\hbar J_q},
	\label{eq:mua2}
	\\
	\frac{E_{\alpha q}^{(s)}}{N} = \frac{\hbar^2q^2}{2M} \mp \nu\frac{|\gamma_q|}{\alpha}.
	\label{eq:Ea2}
\end{align}

Figure~\ref{fig:complex} shows two examples for the same linear coupling $\tilde\nu=0.3$ and interaction parameter $\kappa N=20\pi$: an in-phase, unpolarized state with $\alpha=0.8$, and a polarized state with $\alpha=6$ and $\pi$ relative phase. 
While the former is dynamically stable against small perturbations, as its time evolution~(Fig.~\ref{fig:evol_linear}) shows, the latter is not.

\section{Solitonic states } \label{sec:Solitons}


\begin{figure}[tb]
	\flushleft 	({\bf a})\\
	\includegraphics[width=\linewidth]{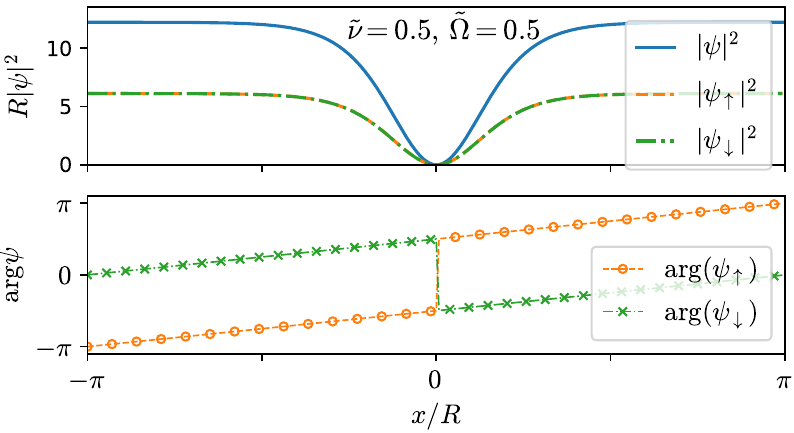}\\ 
	\vspace{-0.4cm}	({\bf b})  
	\includegraphics[width=\linewidth]{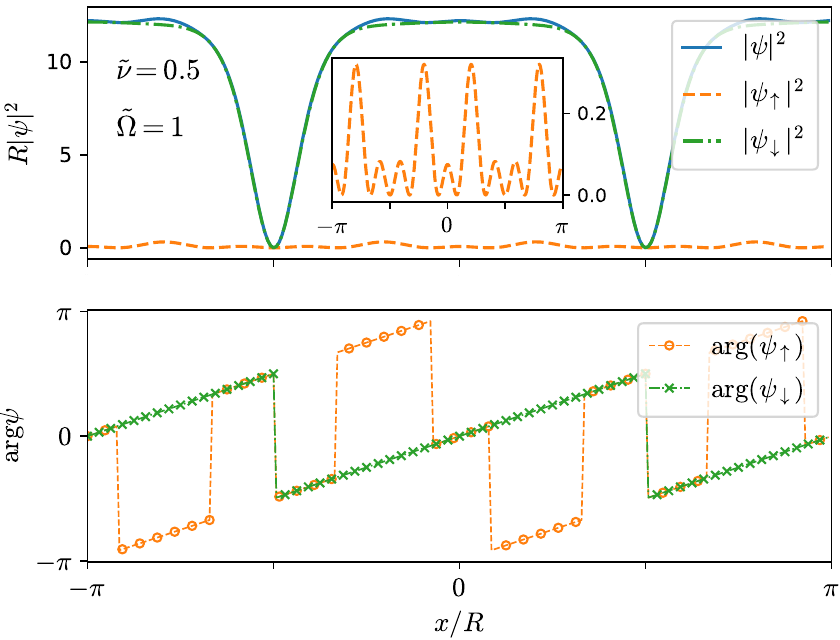}
	\caption{Dark soliton states in chiral spinor condensates with interaction parameter $\kappa N=20\pi$ and linear coupling $\tilde\nu=0.5$. (a) Dark solitons with relative phase $\varphi=\pi$ at $\tilde\Omega=0.5$. (b) Two dark solitons at $\tilde\Omega=1$ in a highly imbalanced state with non-constant relative phase. The inset zooms in the density of the minority component. }
	\label{fig:DSs}
\end{figure}
Apart from plane-wave superpositions, which present a non-homogeneous density profile and constant current densities, one can find generic nonlinear excited states (we will refer to them as solitonic states) that present both density and current space-varying profiles. 
Among the simplest states of this type, we already introduced Eq.~\eqref{eq:dn_general}, which replicates in the spinor system the solutions of the scalar chiral system. 
This is illustrated in Fig.~\ref{fig:DSs}(a) for unpolarized, out-of-phase ($\psi_\uparrow=-\psi_\downarrow$) dark solitons at $\Omega=0.5\,\Omega_0$. 
The usual $\tanh(x/\xi)$ functional form of infinite systems transforms here, within a ring trap, into the Jacobi $\mbox{sn}(x/\xi,\mathfrak{m})$ function, a particular case of Eq.~\eqref{eq:dn_general}~\cite{Arazo2023}. 
The spinor system allows also for more complex structures involving dark solitons. 
Figure \ref{fig:DSs}(b) shows two strongly polarized dark solitons at rotation rate $\Omega=\Omega_0$. 
Interestingly, the minority component presents a highly irregular density profile (see the inset) sustained by sudden $\pi$ jumps in the phase profile that produce alternate regions of either in-phase $\varphi=0$ or out of phase $\varphi=\pi$ spin components.

More generally, as it happens in the presence of contact interparticle interactions, one can distinguish two types of solitonic states in spinor systems: regular (dark or bright) solitons and Josephson vortices~\cite{Kaurov2005}. 
While the former solutions, which do not have Josephson currents, are well known in scalar condensates (see, e.g., \cite{Pitaevskii}), the latter ones are characterized by the presence of Josephson currents and are only present in spinor systems~\cite{Qadir2012,Sophie2018,Baals2018}. 
For repulsive, contact interparticle interactions, dark solitons and Josephson vortices can be considered as domain walls of the total and relative phase with corresponding healing lengths $\xi={\hbar^2/(m\mu)}$ and $\xi_\nu={\hbar^2/(4m\,\nu)}$, respectively~\cite{Son2002}, and the interconversion between them takes place at $\xi=\xi_\nu$~\cite{Kaurov2005,Kaurov2006}. 
For attractive interactions, bright solitons can also support strongly localized Josephson vortices~\cite{Radek2024}. 
As we show below, when current-density interactions are acting, one can find all of these types of stationary solitonic states. In general, however, their stability is diminished or lost in the presence of spin currents.

\begin{figure}[tb]
	\flushleft 	({\bf a})\\
	\includegraphics[width=\linewidth]{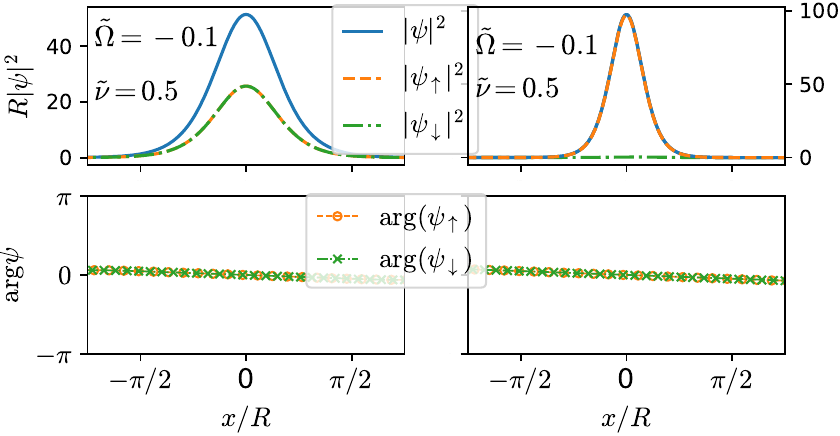}\\ 
	\vspace{-0.5cm}
	\flushleft 	({\bf b})\\
	\includegraphics[width=\linewidth]{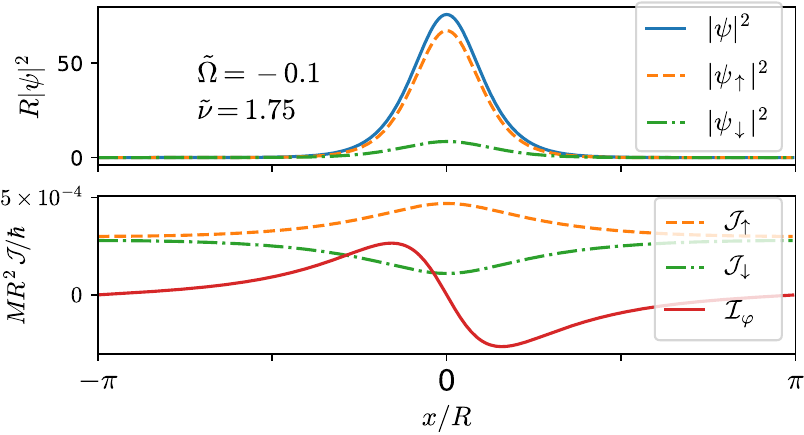}	
	\caption{Spinor bright-soliton states in a system with interaction parameter $\kappa N=20\pi$ and rotation rate $\tilde\Omega=-0.1$. (a) Unpolarized (left) and strongly polarized (right) solitons for the same linear coupling $\tilde\nu=0.5$. (b) Density (top) and current (bottom) profiles of a polarized soliton at linear coupling $\tilde\nu=1.75$. }
	\label{fig:BS}
\end{figure}
Starting with bright solitons, Fig.~\ref{fig:BS} shows our numerical results for three stationary states of this type with rotation rate $\tilde{\Omega}=-0.1$. Panels (a) illustrate, for fixed linear coupling, the differences between unpolarized (left) and polarized (right) solitons. 
While the former exhibits a null relative phase, the latter presents tiny spin currents. 
Such currents increase at higher linear coupling, as in the case shown in panel (b) and, along with Josephson currents, give rise to weak loop patterns  centered in the junction between spin components that are analogue to Josephson vortices in systems with repulsive interactions~\cite{Radek2024}.

\begin{figure}[tb]
	\includegraphics[width=\linewidth]{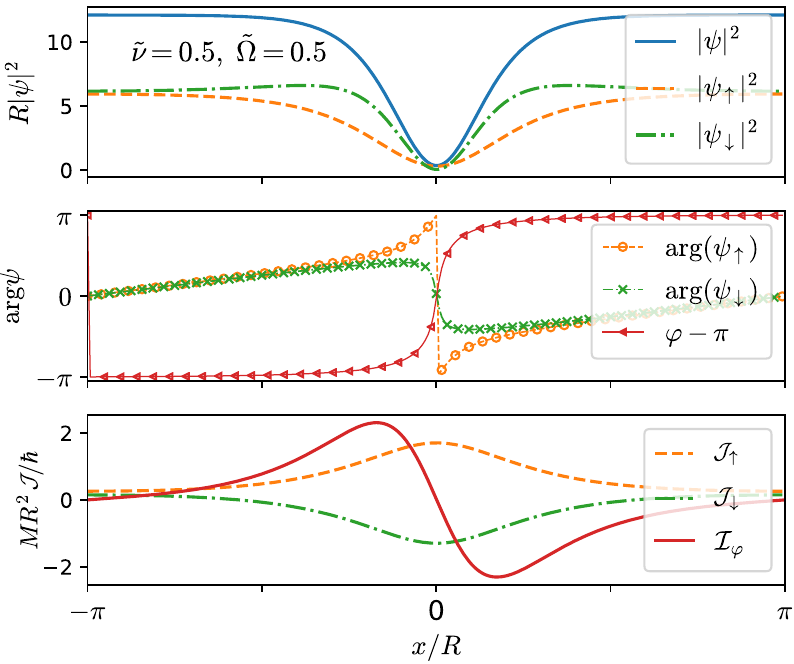}
	\caption{Josephson-vortex state in a chiral spinor condensate. Density profiles (top), phase profiles (middle), and currents (bottom) are shown for a system with interaction parameter $\kappa N=20\pi$, linear coupling $\tilde\nu=0.5$, and rotation rate $\smash{\tilde\Omega}=0.5$. }
	\label{fig:JV}
\end{figure}
The loop currents within bright solitons can be compared with their counterpart at $\tilde \Omega>0$, when the current-density interaction is effectively repulsive. 
Figure~\ref{fig:JV} depicts our numerical results for a stationary Josephson-vortex state in a chiral spinor system with equal parameters as for the dark soliton shown in Fig.~\ref{fig:DSs}(a). 
Although the total density profile resembles the dark soliton, the $2\pi$ jump in the relative phase (central panel) stands out as its main signature, whereas the non-vanishing Josephson current $\mathcal{I}_\varphi$ (bottom panel) changes sign around the vortex core, which is signaled by the density minimum. 
The represented current densities are measured with respect to the rotating frame $\mathcal{J}_\sigma=J_\sigma-n_\sigma\Omega R$, and show opposite directions as measured with respect an average non-zero current. 
An important difference with respect to regular (non-chiral) static Josephson vortices, analytically described in Ref. \cite{Kaurov2005} as $\psi_{\uparrow,\downarrow}\propto \tanh(x/\xi_\nu)\pm i\beta/\cosh(x/\xi_\nu)$, resides in the uneven spin density profiles, which arise from differences in the effective interactions induced by the chiral currents, and are closer to those of regular moving Josephson vortices~\cite{Sophie2018}.

As for multiple Josephson vortices, two configurarions are possible: co- and counter-rotating vortices.
\begin{figure}[tb]
	\includegraphics[width=\linewidth]{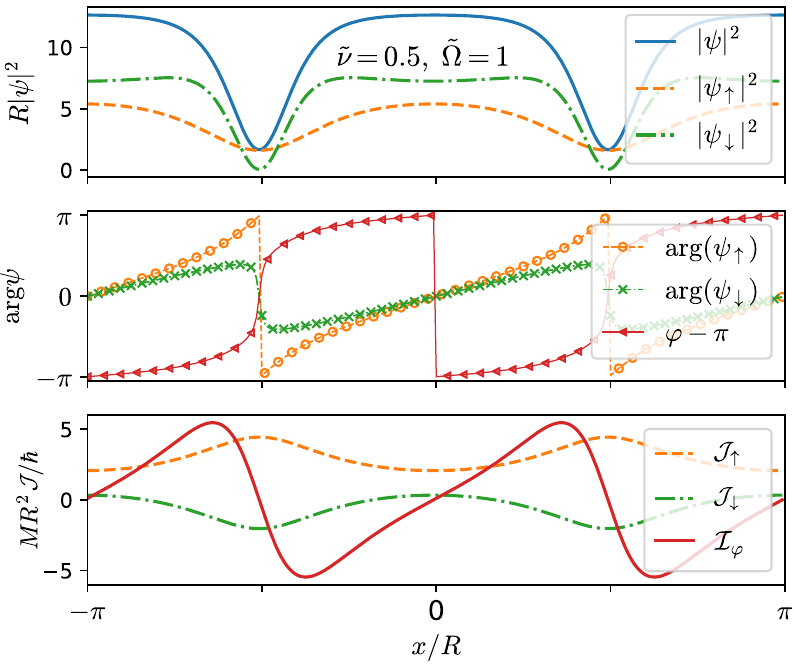}
	\caption{Co-rotating Josephson vortices  with interaction parameter $\kappa N=20\pi$ at $\tilde\Omega=1$ and $\tilde\nu=0.5$. We show the density profile (top), phase (middle) and currents in the rotating frame (bottom). The presence of the Josephson vortices is signaled by a non-zero density and a $2\pi$-phase jump at the core of the vortices. Note that the Josephson current $\mathcal{I}_\varphi$ changes sign at the core of the vortices and also at half distance between them.}
	\label{fig:2JV}
\end{figure}
Figure~\ref{fig:2JV} illustrates the case of a state with two co-rotating Josephson vortices, which are associated with corresponding smooth $2\pi$ jumps in the relative phase (central panel) and a neat non-vanishing density at the vortex cores. Notice that these cores are located in the juction at the $x$-position of the density minima of both spin components, and that the Josephson current changes sign (vanishes) also at half distance between them.
	
\begin{figure}[tb]
	\includegraphics[width=\linewidth]{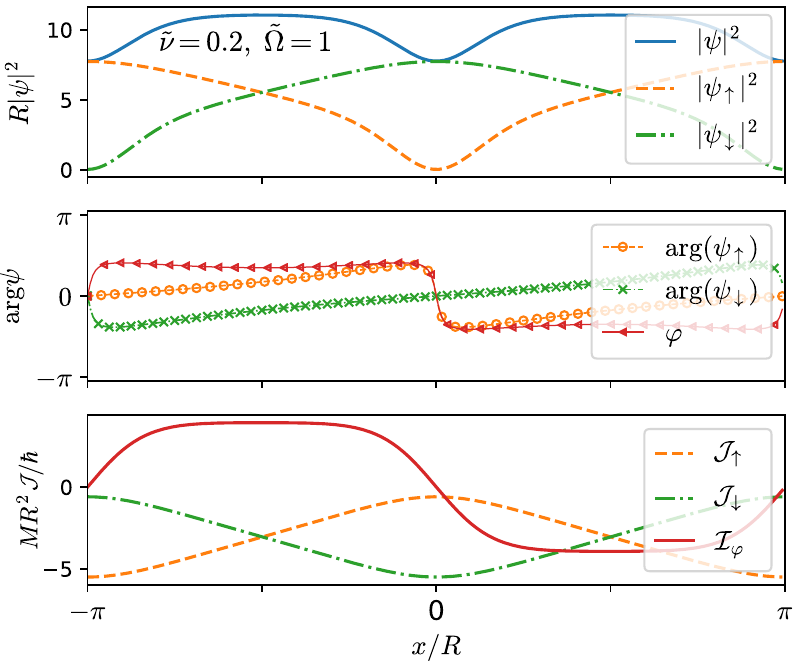}
	\caption{Counter--rotating Josephson vortices  with interaction parameter $\kappa N=20\pi$ at $\tilde\Omega=1$ and $\tilde\nu=0.2$. As in co-rotating Josephson vortices, the component density (top) is non-zero and the Josephson current (bottom) changes sign at the cores of the vortices; the relative phase (middle), however, now shows $\pi$ jumps. }
	\label{fig:2JVb}
\end{figure}
The scenario is more involved for counter-rotating Josephson vortices. 
As shown in Figure \ref{fig:2JVb}, although one can identify the sign change of Josephson currents around the vortex cores (now signaled by minima in the total density) the relative phase experiences just  $\pi$ (opposite) jumps across them. 
The latter are caused by staggered dark-soliton-like phase profiles in the spin components. 
These features are in sharp contrast with the case of static counter-rotating Josephson vortices in regular spinor condensates~\cite{Qiu2021}, and present common features with other moving solitonic structures, as the staggered dark-solitons or Manakov solitons of Ref.~\cite{Sophie2018}.


\begin{figure}[tb]
	\includegraphics[width=\linewidth]{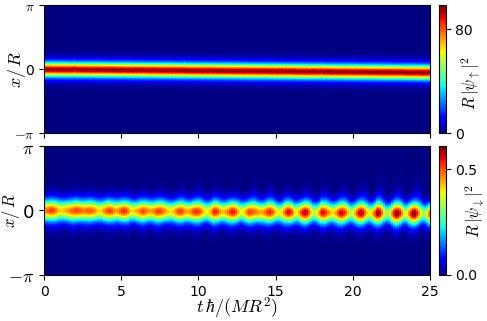}\\
	\includegraphics[width=\linewidth]{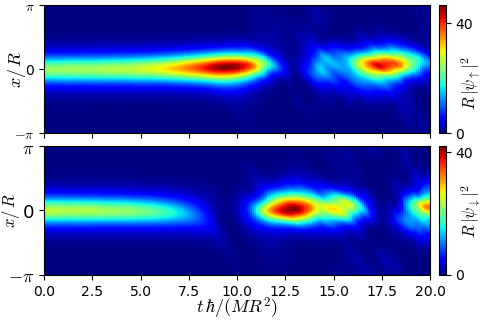}
	\caption{Real-time evolution of the polarized (top) and unpolarized (bottom) bright solitons shown in Fig. \ref{fig:BS}(a). The latter states decay rapidly, while the former ones at least preserve for long time the strongly polarized features of their density profiles due to the small coherent coupling considered.}
	\label{fig:evol_bs}
\end{figure}
\begin{figure}[tb]
	\includegraphics[width=\linewidth]{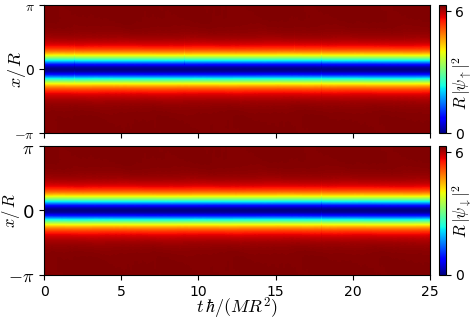} \\
	\includegraphics[width=\linewidth]{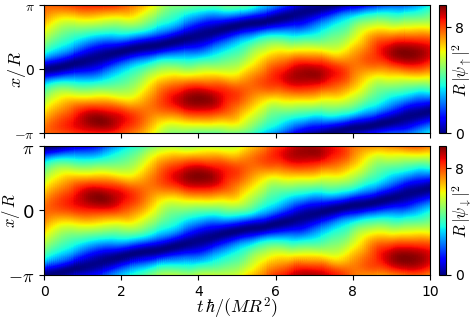}
	\caption{Real-time evolution of the dark solitons and Josephson vortices shown in Fig. \ref{fig:DSs}(a) and \ref{fig:2JVb}. The dark-soliton states are stable, since they do not present Josephson currents. In Josephson-vortex states, on the other hand, the presence of Josephson currents destabilizes the system. }
	\label{fig:evol_ds}
\end{figure}
To test the dynamical stability of these solitonic states, we have performed numerical simulations of the Gross-Pitaevskii Eq.~\eqref{eq:gp} to obtain the real-time evolution after adding sinusoidal perturbations of small amplitude on the stationary states~(as we did for plane-wave superpositions in Fig.~\ref{fig:evol_linear}).
Figure~\ref{fig:evol_bs} illustrates the case of polarized and unpolarized bright solitons, and Fig.~\ref{fig:evol_ds} shows the case of dark solitons and Josephson vortices. 

The time evolution of solitonic states shows, as a common feature, the decay of states with non-zero Josephson currents and, in general, of polarized states. 
However, polarized bright solitons [see right panel of Fig. \ref{fig:BS}(a) and top panel of Fig.~\ref{fig:evol_bs}], which have effective attractive interactions, are at least structurally stable (maintaining their density profiles) at low linear coupling.

	
\section{Discussion and conclusions} \label{sec:Conclusions}

As a generalization of the recently-realized scalar BECs with current-density interactions, we have considered a spin-$1/2$ condensate with intra-spin  current-density interactions. By means of both analytical and numerical methods, within a mean field framework, we have explored the steady and stability properties of plane waves, bright and dark solitons, and Josephson vortices. Our results show the manifest chiral dynamics of these states induced by the interactions. In addition,
 the interplay between the spin current densities and the linear coupling, which allows for spin flips, gives rise to an effective, nonlinear spin-orbit coupling that results in unexpected features:  neither the presence of population imbalance nor the flow of Josephson current between the spin components, which points to differences in the spin currents and thus in the interactions, favor stability. Among the stable states, we have found nonlinear waves made of linear superpositions that replicate the spectrum of linear spinor systems. These findings are clearly within reach of current experimental research. 
 
Although, in order to focus on the effect of current density interactions, we have not considered additional contact interactions, recent experiments do include both types of interactions \cite{Frolian2022}. The emergence of our model from this scenario, in a crossover with varying ratio between the strength of both interactions, and from the direct 2D analysis of realistic systems, is left for a future work.


\begin{acknowledgments}
	We are grateful to Alessio Celi for insightful discussions.
	This work has been funded by Grant No.~PID2020-114626GB-I00 from the MICIN/AEI/10.13039/501100011033 and by the European Union Regional Development Fund within the ERDF Operational Program of Catalunya (project QuantumCat, Ref. No. 001-P001644).
	V. D. acknowledges support from Agencia Estatal de Investigaci\'on (Ministerio de Ciencia e Innovaci\'on, Spain) and Fondo Europeo de Desarrollo Regional (FEDER, EU) under grant PID2019-105225GB-100.
\end{acknowledgments}

\bibliography{two_rings}

\end{document}